\documentclass[prb,preprint,amsmath,amssymb,superscriptaddress,longbibliography]{revtex4-1}

\usepackage[utf8]{inputenc}
\usepackage[english]{babel}
\usepackage[babel,german=quotes]{csquotes}
\usepackage[T1]{fontenc}
\usepackage{amsmath}
\usepackage{amssymb}
\usepackage{graphicx}
\usepackage{hyperref}

\usepackage{soul}
\usepackage{braket}

\usepackage{xcolor}

\definecolor{mlgreen}{rgb}{.035,.6,.251}
\definecolor{mlviolett}{rgb}{.643,.259,.804}

\begin{document}

\title{Quantum susceptibilities in time-domain sampling of electric field fluctuations}
\author{Matthias Kizmann}
\affiliation{Department of Physics and Center for Applied Photonics,
University of Konstanz, D-78457 Konstanz, Germany}
 \author{Andrey S. Moskalenko}
 \email{moskalenko@kaist.ac.kr}
 \affiliation{Department of Physics, KAIST, Daejeon 34141, Republic of Korea}
 \affiliation{Department of Physics and Center for Applied Photonics,
University of Konstanz, D-78457 Konstanz, Germany}
 \author{Alfred Leitenstorfer}
 \affiliation{Department of Physics and Center for Applied Photonics,
University of Konstanz, D-78457 Konstanz, Germany}
  \author{Guido Burkard}
  \affiliation{Department of Physics and Center for Applied Photonics,
University of Konstanz, D-78457 Konstanz, Germany}
 \author{Shaul Mukamel}
\email{smukamel@uci.edu}
  \affiliation{Department of Chemistry and Physics and Astronomy, University of California, Irvine, California 92697-2025, USA}

\begin{abstract}
Electro-optic sampling has emerged as a new quantum technique enabling measurements of electric field fluctuations on subcycle time scales. Probing a second-order nonlinear material with an ultrashort coherent laser pulse imprints the fluctuations of a terahertz field onto the resulting near-infrared electro-optic signal. We describe how the statistics of this time-domain signal can be calculated theoretically, incorporating from the onset the quantum nature of the electric fields involved in the underlying interactions. To this end, a microscopic quantum theory of the electro-optic process is developed using an ensemble of non-interacting three-level systems as a model for the nonlinear material. We find that the response of the nonlinear medium can be separated into a classical part sampling the terahertz field and quantum contributions independent of the state of the probed terahertz field. The quantum response is caused by interactions between the three-level systems mediated by the terahertz vacuum fluctuations. It arises due to cascading processes and contributions described by quantum susceptibilities solely accessible via quantum light. We show that the quantum contributions can be substantial and might even dominate the total response. We also determine the conditions under which the classical response serves as a good approximation of the electro-optic process and demonstrate how the statistics of the sampled terahertz field can be reconstructed from the statistics of the electro-optic signal. In a complementary regime, electro-optic sampling can serve as a spectroscopic tool to study the pure quantum susceptibilities of materials.
\end{abstract}
\maketitle

Nonlinear optics with laser light serves as one of the fundamental tools in modern experimental physics. In three-wave mixing spectroscopy for example, materials are examined by irradiating them with optical fields and measuring the light emitted into a new direction. The spectral and temporal shape as well as the amplitude and phase of the incoming fields can be varied for the study of different effects.  These options lead to a wide range of classical spectroscopic techniques such as coherent  anti-Stokes Raman\cite{Tolles1977}, photon echo\cite{Cho1992}, two-dimensional femtosecond spectroscopy\cite{Jonas2003} and many more. The employment of strong coherent states of laser light justifies a classical treatment of the electric fields. The corresponding processes inside the matter are then described by a classical response in terms of classical causal nonlinear susceptibilities \cite{Mukamel1995,Boyd_book,scully_1997}. We denote this scenario involving quantum matter and classical fields as classical nonlinear optics. If quantum fields are involved\cite{Mukamel2020} on the other hand, some field degrees of freedom are described by nonclassical states of light where phenomena such as superposition\cite{Kira2011} and entanglement\cite{Yabushita2004,Kalashnikov2014} can be exploited to induce useful effects in the nonlinear medium. Here, the material response is not necessarily described by classical susceptibilities. Nonclassical states of light can be generated by nonlinear processes such as four-wave mixing\cite{Silberhorn2007} or parametric downconversion\cite{Kwiat1999,Mosley2008,Sharapova2015}.

Usually, the nonlinear susceptibilities are calculated separately from the involved electric fields and then inserted into an effective Hamiltonian for the nonlinear interaction\cite{Boyd_book,scully_1997}. This derivation of the nonlinear susceptibility assumes that all involved electric fields are left unaltered by the interaction with matter which is true for classical strong coherent fields which commute and correlations between these fields can be neglected. In quantum optics, these correlations are the key characteristics of the quantum fields. Here, the particular time ordering of the interactions between the fields and matter play a crucial role since the fields are also affected by these interactions in the course of the nonlinear process. This scenario is most adequately described by quantum susceptibilities that are influenced by changes in higher-order fluctuations of the nonlinear medium\cite{Roslyak2009,Dorfman2012,Dorfman2016}.

In its standard application, electro-optic sampling represents a typical example for classical nonlinear optics. Here, a short near-infrared probe pulse is used to sample the trace of a classical terahertz (THz) field with subcycle temporal resolution\cite{Wu1995,Leitenstorfer1999,Keiber2016}. The fields interact in a second-order $\chi^{(2)}$ nonlinear crystal and the polarization of the near-infrared probe changes proportionally to the local THz field amplitude. This technique was recently extended into the quantum domain by applying it to quantum THz fields with vanishing mean values (e.g. vacuum fluctuations). The probe is sampling the fluctutations of the THz field on subcycle time scales and the fluctuations of the change in the polarization of the probe are related to those of the sampled THz field\cite{Riek2015,Benea2019}. The striking advantage of electro-optic sampling in comparison to standard homodyne detection\cite{Raymer1995} is that it provides a direct time-resolved probe of the fluctuations of THz fields (analogous to the sampling of a classical THz field)\cite{Riek2017,Guedes2019,Kizmann2019}. A slightly modified electro-optic configuration also allows for a characterization of the frequency-resolved fluctuations of THz fields\cite{Benea2016,Benea2019} and might provide a useful spectroscopic tool to measure, e.g., the linear dielectric function of materials without requiring any incoming photons in the probed frequency range\cite{Liberato2019}. Theoretical models have so far relied on a macroscopic description of the second-order nonlinear interaction inside the crystal, where a frequency-independent classical second-order susceptibility $\chi^{(2)}$ is used to model the underlying nonlinear process\cite{Moskalenko2015,Lindel2020} in the electro-optic medium. Here, the sampling of a classical THz field is straightforwardly extended to the sampling of quantum fields. We investigate the limitations of this approach and show when quantum effects of the nonlinear response must be taken into account.

We present a time-domain quantum electrodynamic theory of electro-optic sampling of the quantum properties of the THz electric field. A system of 3-level noninteracting molecules is employed to model the nonlinear medium and to calculate its frequency response in amplitude and phase. We demonstrate how fluctuations of the phase-dependent quadratures of the generated electro-optic signal can be measured and that a full quantum tomography of the generated signal is feasible. Our theory focuses on the nature of the nonlinear electro-optic response and pinpoints under which conditions a straightforward reconstruction of the quantum fluctuations of the sampled field is possible.

Electro-optic sampling is usually performed with nonlinear crystals in the off-resonance configuration, where the frequencies of all involved fields are far from the resonances providing the second-order nonlinear response. We show that the fluctuations imprinted on the polarization of the probe field are not solely given by the classical response of the nonlinear medium, but also contain quantum corrections originating from the quantum THz field. We identify two types of quantum corrections: quantum susceptibilities and cascading processes which lead to an effective intermolecular interaction mediated by THz fluctuations.
These corrections are independent on the actual state of the sampled THz field. Consequently, the classical response of the nonlinear medium alone provides information about the sampled THz field. Previous theories of electro-optic sampling can describe the cascading processes macroscopically by relying on the Maxwell equations. However, the microscopic treatment reveals additional contributions that lead to squeezing of the generated near-infrared field for certain phase shifts. The contributions due to the quantum susceptibilities are neglected entirely by the macroscopic description. We identify conditions under which electro-optic sampling might be employed as a spectroscopic tool to study these quantum susceptibilities. Finally, we describe how the probability distribution of the bare THz fluctuations can be reconstructed from the measured probability distribution of the electro-optic signal. We find that quantum corrections can affect this reconstruction.

\section{THE SETUP}
\begin{figure}[!b]
\includegraphics[width=\columnwidth]{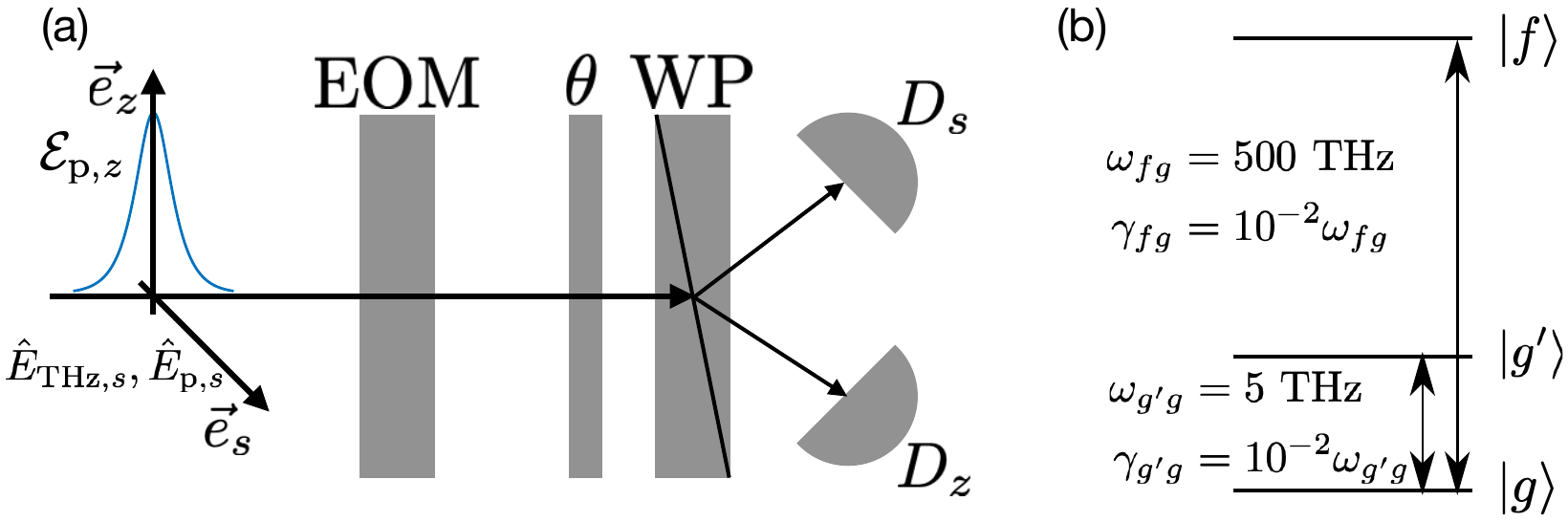}
\caption{Setup for electro-optic sampling and model level scheme of the molecules. (a), setup consisting of a nonlinear electro-optic medium (EOM), a waveplate inducing a phase shift of $\theta$ along its fast axis, a Wollaston prism (WP) separating the two polarization directions and two photon detectors $D_s$ and $D_z$. An intense $\vec{e}_z$-polarized coherent near-infrared (NIR) probe pulse (blue) $\mathcal{E}_{\mathrm{p},z}$ is applied to the setup to probe a co-propagating quantum THz field $\hat{E}_{\mathrm{THz},s}$ polarized in the perpendicular $\vec{e}_s$-direction. $\hat{E}_{\mathrm{p},s}$  denotes the induced $\vec{e}_s$-polarized NIR quantum field. (b), level scheme for the molecules representing the EOM, comprising 3 levels $i,j=g,g',f$ with transition frequencies $\omega_{ij}$ and lifetime broadenings $\gamma_{ij}$. \label{Fig1}}
\end{figure}

The setup geometry and the model level scheme of the medium subjected to the electro-optic detection are sketched in Figs.~\ref{Fig1}(a) and (b), respectively.
We assume an effective three-level electro-optic medium (EOM), depicted in Fig.~\ref{Fig1}(b), that possesses a second-order nonlinear susceptibility tensor with a zincblende-type symmetry. We model the EOM by a large number of independent three-level quantum systems which we term \textit{molecules}. An intense near-infrared (NIR) probe field in a strong multimode coherent state $|\{\mathcal{E}_{\mathrm{p},z}\}\rangle$ with a large amplitude $\mathcal{E}_{\mathrm{p},z}(\omega)=\Braket{\{\mathcal{E}_{\mathrm{p},z}\}\left|\hat{E}_{\mathrm{p},z}(\omega)\right|\{\mathcal{E}_{\mathrm{p},z}\}}$ and a terahertz (THz) field $\hat{E}_{\mathrm{THz},s}(\Omega)$ are sent into the EOM. These fields are linearly polarized along the $\vec{e}_z$ and $\vec{e}_s$ directions, respectively.  While propagating through the nonlinear medium, the two fields undergo a sum-frequency generation (SFG) and a difference-frequency generation (DFG) process. Both three-wave mixing processes generate a $\vec{e}_\mathrm{s}$-polarized weak contribution to the NIR field $\hat{E}_{\mathrm{p},s}$. A waveplate then shifts the two polarization components of the NIR field after the EOM by the angle $\theta$ with respect to each other, (see Appendix \hyperref[AppA]{A}), effectively mixing the two beams. The light is then sent through a Wollaston prism which separates the two polarization directions. Finally, the number of photons in these beams, $\hat{N}^\prime_s$ and $\hat{N}^\prime_z$, are measured with the detectors $D_s$ and $D_z$, respectively. The electro-optic signal is defined by their difference,
\begin{align}
\hat{\mathcal{S}}(\theta)=\hat{N}^\prime_z-\hat{N}^\prime_s=C\int_0^\infty\!\!\! \mathrm{d}\omega\; \frac{1}{\hbar\omega}\left[P(\theta)\hat{E}^\dagger_{\mathrm{p},z}(\omega)\hat{E}_{\mathrm{p},s}(\omega)+H.c.\right],\label{signal}
\end{align}
where $C=4\pi\varepsilon_0 A c_0$ with $A$ being the effective transversal area determined by the beam waist of the probe field, $c_0$ the speed of light in vacuum, $\varepsilon_0$ the vacuum permittivity and $P(\theta)=\sqrt{-\cos\theta}+i\sqrt{1+\cos\theta}$. An angle of $\theta=\pi/2$ ($\theta=\pi$) is produced by a quarter-wave (half-wave) plate\cite{Sulzer2020} [see Fig.~\ref{Fig1}(a)].

The mean value of the electro-optic signal $\hat{\mathcal{S}}(\theta)$ is proportional to that of the sampled THz field\cite{Wu1995}. The temporal profile of a classical THz field can therefore be sampled by measuring the mean value of $\hat{\mathcal{S}}(\theta)$ for different delay times between the probe and the THz field. To achieve subcycle temporal resolution, the NIR probe field must be shorter than the characteristic period of the sampled THz field. For a nonclassical THz field higher moments of the electric field are of interest. To access them experimentally, the statistics of $\hat{\mathcal{S}}(\theta)$  needs to be gathered. The relative frequency of occurrence of each measurement outcome $\mathcal{S}$ is collected to build a histogram which samples the probability distribution $P(\mathcal{S},\theta)$ equivalently to standard balanced homodyne tomography. Theoretically, $P(\mathcal{S},\theta)$ is given by (see Appendix \hyperref[AppA]{A})
\begin{align}
\begin{split}
P(\mathcal{S},\theta)&=\left\langle :\! \frac{1}{\sqrt{2\pi N}}e^{-\frac{\left(\mathcal{S}-\hat{\mathcal{S}}(\theta)\right)^2}{2N}}\!:\right\rangle,\\
&=\frac{1}{\sqrt{2\pi N}}\sum_{k=0}^\infty \frac{1}{(2N)^{k/2} k!}H_k(\mathcal{S}/\sqrt{2N})\exp\left(-\frac{\mathcal{S}^2}{2N}\right)\langle :\! \hat{\mathcal{S}}^k(\theta)\! :\rangle,
\end{split}\label{prob}
\end{align}
where $N=C\int_0^\infty \mathrm{d}\omega \left|\mathcal{E}_{\mathrm{p},z}(\omega)\right|^2/\hbar\omega$ is the mean number of photons of the probe, $H_k(x)$ is the $k$th-order Hermite polynomial and the colons denote normal ordering.

\section{SUPEROPERATOR REPRESENTATION OF CLASSICAL AND QUANTUM NONLINEAR SUSCEPTIBILITIES}
In this work we want to ignore corrections to the probability distribution related to the mean value of the signal $\langle:\!\hat{\mathcal{S}}(\theta)\!:\rangle$ since they only provide information about the classical characteristics of the sampled THz field, and focus on the normally-ordered second moment of the signal, $\Gamma=\langle :\! \hat{\mathcal{S}}^2(\theta)\! :\rangle$. The variance of $\hat{\mathcal{S}}(\theta)$ is given by $\langle\hat{\mathcal{S}}^2(\theta)\rangle=N+\Gamma$, where $N$ is the shot noise of the probe. We thus focus on nonclassical THz fields with a vanishing electro-optic expectation value $\langle\hat{\mathcal{S}}(\theta)\rangle$ and study the influence of the fluctuations of these THz fields on $\Gamma$. Moreover, we consider the sampling of THz vacuum fluctuations since the vacuum represents a natural experimental resource and the physics remains maximally transparent in this case (an expression for $\Gamma$ that describes the sampling of arbitrary quantum fields is derived in Appendix \hyperref[AppB]{B}). We employ the superoperator formalism\cite{Mukamel1995} which offers a compact quantum treatment of both the matter system and the electric fields to derive a microscopic expression for $\Gamma$. Within this formalism, with any ordinary operator $\hat{A}$ we associate plus and minus type superoperators defined by their action on an arbitrary operator $\hat{X}$;  $\hat{A}_+\hat{X}=\frac{1}{2}\{\hat{A},\hat{X}\}=\frac{1}{2}\left(\hat{A}\hat{X}+\hat{X}\hat{A}\right)$ and $\hat{A}_-\hat{X}=[\hat{A},\hat{X}]=\hat{A}\hat{X}-\hat{A}\hat{X}$. The evolution of the system is determined by the dipole light-matter interaction Hamiltonian  $H_\mathrm{\mathrm{int}}(t)=-\sum_{\alpha=z,s}\int_V\mathrm{d}^3r~\hat{\varepsilon}_\alpha(\vec{r},t)\hat{\mathcal{V}}_{\alpha}(t)$, where $V$ denotes the volume of the electro-optic medium. The corresponding superoperator $\hat{H}_\mathrm{\mathrm{int},-}(t)$ is given by\cite{Roslyak2009}
\begin{align}
\begin{split}
\hat{H}_\mathrm{\mathrm{int},-}(t)&=-\sum_{\alpha=z,s}\int_V\mathrm{d}^3r~\hat{\varepsilon}_{\alpha,+}(\vec{r},t)\hat{\mathcal{V}}_{\alpha,-}(t)+\hat{\varepsilon}_{\alpha,-}(\vec{r},t)\hat{\mathcal{V}}_{\alpha,+}(t),\label{InteractionH}
\end{split}
\end{align}
where we have used the fact that the dipole and the electric field operators commute. Here, $\alpha=z,s$ represents the two possible mutually perpendicular polarizations, $\hat{\varepsilon}_\alpha(\vec{r},t)=\hat{E}_{\mathrm{p},\alpha}(\vec{r},t)+\hat{E}_{\mathrm{THz},\alpha}(\vec{r},t)$ is the sum of all relevant field modes and $\hat{\mathcal{V}}_\alpha(t)=\hat{V}_\alpha(t)+\hat{V}^\dagger_\alpha(t)$ denotes the dipole operator in the interaction picture with $\hat{V}_\alpha=\mu_{\alpha,\mathrm{gg'}}\Ket{g}\!\Bra{g'}+\mu_{\alpha,\mathrm{gf}}\Ket{g}\!\Bra{f}+\mu_{\alpha,\mathrm{g'f}}\Ket{g'}\!\Bra{f}$ and $\mu_{\alpha,ij}$ as the dipole moment for the $j\rightarrow i$ transition ($i,j=g,g',f$).

The normally-ordered second moment of the electro-optic signal $\Gamma$ is calculated from the time-dependent density matrix of the entire system of field and matter $\hat{\rho}$,
\begin{align}
\label{Gamma}
\begin{split}
\Gamma&\equiv\langle :\! \hat{\mathcal{S}}^2(\theta)\! :\rangle=\mathrm{tr}\left\{:\!\hat{\mathcal{S}}^2(\theta)\!:\mathcal{T}\exp\left(-\frac{i}{\hbar}\int_{-\infty}^\infty\mathrm{d}t\hat{H}_{\mathrm{int},-}(t)\right)\hat{\rho}_\mathrm{in}\right\}.
\end{split}
\end{align}
Here, $\hat{\rho}_\mathrm{in}=\hat{\rho}_\mathrm{field}\otimes\hat{\rho}_\mathrm{mat}$ denotes the initial density matrix, given by a direct product of the field and matter density matrices and $\mathcal{T}$ represents the time-ordering operator for the superoperators. We assume $\hat{\rho}_\mathrm{field}=\Ket{\{\mathcal{E}_{\mathrm{p},z}\}}\!\Bra{\{\mathcal{E}_{\mathrm{p},z}\}}\otimes \Ket{0_\mathrm{s}}\!\Bra{0_\mathrm{s}}$, i.e. the electric field consists of a multimode coherent state in the $\vec{e}_z$-polarized NIR range and the electromagnetic vacuum in both the $\vec{e}_s$-polarized NIR and THz ranges. The matter system consists of an ensemble of noninteracting molecules initially in the ground state. Since the trace operation is invariant under cyclic permutation, we can let the time evolution according to the interaction Hamiltonian in Eq.~\eqref{InteractionH} act on $:\!\hat{\mathcal{S}}^2(\theta)\!:$. The trace can then be factorized into a product of traces over the field and the matter degrees of freedom. The matter trace is given by a time-ordered product of $n$ Green's functions of superoperators for the $(n+1)$th-order perturbation term in $\hat{H}_{\mathrm{int},-}(t)$, resulting in the $n$th-order susceptibility.

We define the second-order susceptibility
\begin{align}
 \begin{split}
 \chi^{(2)}_{+rs}\left(-(\omega_2+\omega_1);\omega_2,\omega_1\right)=&\frac{1}{\varepsilon_0\hbar^2}\int_{0}^\infty\!\!\!\!\int_{0}^\infty\!\!\!\!\mathrm{d}t_2\mathrm{d}t_1 e^{i\omega_2 t_1}e^{i\omega_1 (t_2+t_1)}\\
&\qquad\times \mathrm{Tr}\left\{\hat{\mathcal{V}}_+(0)\hat{\mathcal{V}}_{r}(-t_1)\hat{\mathcal{V}}_{s}(-\left(t_2+t_1)\right)\hat{\rho}_\mathrm{mat}\right\},\label{susc}
\end{split}
 \end{align}
 where $r,s=\pm$ indicate the type of superoperator dipole operators, and the subscript $(+rs)$ denotes the sequence of the time-ordered superoperators of the dipole operators $\hat{\mathcal{V}}(t)$ (an explicit formula for the susceptibilities is given in Appendix \hyperref[AppB]{B}). Note that the last interaction with the dipole operator has to be of the $'+'$ type since the expectation value in Eq.~\eqref{susc} would otherwise be given by a trace over a commutator, which vanishes due to the invariance of the trace under cyclic permutation.  Earlier interactions can be of either the $'+'$ or the $'-'$ type and are accompanied by the opposite superoperator interaction for the corresponding field [see Eq.~\eqref{InteractionH}]. The various sequences of superoperator interactions can be used to differentiate between classes of second-order susceptibilities.

Usually, the second-order susceptibility (as well as higher-order susceptibilities) is calculated under the assumption that the involved electric fields are classical. This leads to a classical susceptibility of the form $\chi^{(2)}_{+--}$, i.e.  $n$ $'-'$-type dipole operator interactions, followed by a $'+'$-type interaction for the $n$th-order susceptibility. Accordingly, the sequence of superoperator interactions for the fields is given by its conjugate $(-++)$. These types of susceptibilities are also denoted causal since the matter system interacts with two classical fields (first two $'+'$-type interactions) which are left unaltered by the interaction and as a result generates a new field (last $'-'$-type interaction). Another class of second-order susceptibilities is given by $\chi^{(2)}_{++-}$, where the corresponding sequence of superoperator interactions for the electric fields is $(--+)$. Here, the first interaction of the medium with the field leaves the field unaltered, while both the second and the third interactions affect the state of the respective fields. This susceptibility as well as any kind of susceptibilities with more than one $'-'$-type interaction with the fields are called quantum or noncausal susceptibilities since matter and field mutually affect each other and they only appear if a nonclassical electric field takes part in the process\cite{Dorfman2016}.

\section{NORMALLY-ORDERED SECOND MOMENT OF THE THz VACUUM FIELD}

\begin{figure}[ht!]
\includegraphics[scale=2]{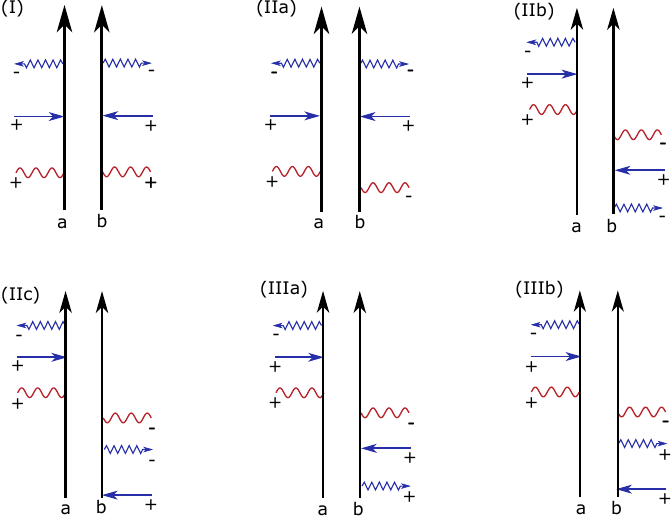}
\caption{Diagrams showing the leading contributions to the electro-optic sampling of the THz vacuum fluctuations [Eq.~\eqref{result}] for two molecules $a$ and $b$. Each diagram depicts the time evolution of both the electric field modes and the density matrix of the matter system. The vertical arrows indicate the time evolution of molecules $a$ and $b$ from the past (bottom) to the present (top), respectively. The red wavy lines denote an interaction with the THz vacuum field (either $\hat{E}_{\mathrm{THz},s}$ or $\hat{E}^\dagger_{\mathrm{THz},s}$), the blue zigzag arrows pointing to the left (right) represent interactions with the  detected NIR field $\hat{E}^\dagger_{\mathrm{p},s}$ ($\hat{E}_{\mathrm{p},s}$) and the straight blue arrows pointing to the left (right) denote interactions with the coherent probe field $\mathcal{E}^*_{\mathrm{p},z}$ ($\mathcal{E}_{\mathrm{p},z}$), respectively. The $\pm$ signs next to the horizontal arrows denote the type of superoperator interaction for the corresponding field mode. ($\mathrm{I}$), a process that can  be described by classical susceptibilities. This diagram is the only one captured by the classical treatment. ($\mathrm{II}$a)$-$($\mathrm{II}$c), processes that involve quantum susceptibilities due to two $'-'$-type interactions of molecule $b$ with $\hat{E}_{\mathrm{p},s}$ and $\hat{E}_{\mathrm{THz},s}$. ($\mathrm{III}$a) and ($\mathrm{III}$b), cascading processes that can be described by classical susceptibilities but are not captured by the classical treatment described in the main text. The depicted diagrams survive the rotating wave approximation (RWA). For a detailed calculation and the full set of diagrams see Appendix \hyperref[AppB]{B}.\label{Fig2}}
\end{figure}

In this section, we calculate the normally-ordered second moment $\Gamma$. Since both the NIR $\hat{E}_{\mathrm{p},s}$ and the THz $\hat{E}_{\mathrm{THz},s}$ field modes are initially in the vacuum state, each of them must act at least twice to give a non-vanishing contribution to Eq.~\eqref{Gamma}. To obtain a non-vanishing contribution from the THz vacuum field $\hat{E}_{\mathrm{THz},s}$ in terms of two second-order processes, we need to have two interactions with $\hat{E}_{\mathrm{p},z}$, $\hat{E}_{\mathrm{p},s}$ and $\hat{E}_{\mathrm{THz},s}$, each. Therefore, we have to expand the exponential in Eq.~\eqref{Gamma} to sixth order in $\hat{H}_{\mathrm{int},-}$. We neglect all linear contributions, since we are interested in fluctuations of the electro-optic signal which should not be influenced by $\chi^{(1)}$ processes, provided these fluctuations are detected outside the nonlinear material. Due to the large number of molecules, the dominant contribution to the signal comes from pairs of molecules each interacting three times with the dipole operator; we are thus looking at two second-order processes instead of a single fifth-order process. We will  treat the probe field as classical because of its strong coherent amplitude, i.e. we only take into account interactions of the form $\hat{E}_{\mathrm{p},z,+}$ ($\hat{E}_{\mathrm{p},z,-}$ vanishes for a classical field). This does not apply to the quantum fields $\hat{E}_{\mathrm{p},s}$ and $\hat{E}_{\mathrm{THz},s}$. The leading diagrams for $\Gamma$ that survive the rotating wave approximation (RWA) are given in Fig.~\ref{Fig2}. However, the RWA has not been applied in our calculations, which are based on Eq.~\eqref{InteractionH} and include the full set of diagrams given in Appendix \hyperref[AppB]{B}. In Figure~\ref{Fig2}$(\mathrm{I})$ the pair of molecules that generate the signal interact with the THz vacuum field through $\hat{E}_{\mathrm{THz},s,+}$. In this case, the time ordering of these two interactions is immaterial and the nonlinear processes occurring in the two molecules are completely independent of each other. In contrast, the remaining diagrams describe shared fluctuations between two molecules that involve the commutator of two THz field modes. The first interaction with the THz field $\hat{E}_{\mathrm{THz},s,-}$ on molecule $b$ e.g. emits a THz photon, changing the state of the THz field, which is then picked up by molecule $a$ through the interaction with $\hat{E}_{\mathrm{THz},s,+}$. Therefore, the intermolecular time ordering of these interactions with the THz field does matter. This fact leads to an effective interaction between molecules $a$ and $b$ that is mediated by the THz vacuum fluctuations. These processes can only be understood in the joint space of both molecules. We split the normally-ordered second moment $\Gamma$ of the electro-optic signal into three contributions
\begin{align}
\Gamma&=\Gamma_\mathrm{I}+\Gamma_\mathrm{II}+\Gamma_\mathrm{III}.\label{result}
\end{align}
These will be discussed in detail below.

The first contribution $\Gamma_\mathrm{I}$ is given by
\begin{align}
\Gamma_\mathrm{I}&=\left(\frac{N\omega_\mathsf{p}L}{c_0 }\right)^2\int_0^\infty \!\!\!d\Omega\frac{\hbar\Omega}{C}\left|D(\Omega,\theta)\right|^2,\label{gammaI}
\end{align}
where $\omega_\mathrm{p}=\int_0^\infty\mathrm{d}\omega \left|\mathcal{E}_{\mathrm{p},z}(\omega)\right|^2/\int_0^\infty\mathrm{d}\omega \left(1/\omega\right)\left|\mathcal{E}_{\mathrm{p},z}(\omega)\right|^2$ is the average detected frequency and $L$ is the length of the nonlinear medium along the propagation direction. Hereafter, $\omega$ and $\Omega$ denote frequencies in the NIR and THz range, respectively. We have further introduced the detection window $D(\Omega,\theta)$, which depends on the classical susceptibilities $\chi^{(2)}_{+--}$
\begin{align}
\begin{split}
D(\Omega,\theta)&=\frac{1}{2}\int_0^\infty\mathrm{d}\omega f_-(\omega,\Omega,\theta)\left(\chi^{(2)}_{+--}(-\omega;\Omega,\omega-\Omega)+\chi^{(2)}_{+--}(-\omega;\omega-\Omega,\Omega)\right)\\
&\qquad -\frac{1}{2}\int_0^\infty\mathrm{d}\omega f^*_+(\omega,\Omega,\theta)\left(\chi^{(2)*}_{+--}(-\omega;-\Omega,\omega+\Omega)+\chi^{(2)*}_{+--}(-\omega;\omega+\Omega,-\Omega)\right),\label{class}
\end{split}
\end{align}
with the autocorrelation functions $f_\pm(\omega,\Omega,\theta)=P(\theta)\mathcal{E}^*_{\mathrm{p},z}(\omega)\mathcal{E}_{\mathrm{p},z}(\omega\pm\Omega)/\int_0^\infty\!\mathrm{d}\omega \left|\mathcal{E}_{\mathrm{p},z}(\omega)\right|^2$. The leading diagram to this contribution is depicted in Fig.~\ref{Fig2}($\mathrm{I}$). Here, both molecules interact independently with the THz vacuum since the interaction with $\hat{E}_{\mathrm{THz},s,+}$ does not alter the vacuum state and the matter response can be described in the single-molecule space. In that sense, the THz vacuum may be treated here as a classical field, which is why the matter response is given by classical suscpebtibilities $\chi^{(2)}_{+--}$ [cf. Eq.~\eqref{susc}]. The same result is also obtained by calculating the third-order correction to the electro-optic signal $\hat{\mathcal{S}}(\theta)$ in Eq.~\eqref{signal} and squaring it [see the result for arbitrary THz fields in Eq.~\eqref{Gammaarbf} of Appendix \hyperref[AppB]{B}]. This contribution is always positive. $\Gamma_\mathrm{I}$ represents the classical contribution to the normally-ordered second moment $\Gamma$ and this result may also be obtained by means of an effective second-order nonlinear polarization $\hat{P}^{(2)}$, as was done in previous works\cite{Moskalenko2015,Lindel2020}. This is the only contribution that depends on the state of the THz field and is actually sampling its fluctuations [cf. Eq.~\eqref{Gammaarbf} in Appendix \hyperref[AppB]{B}]. We thus have a quantum extension of the classical electro-optic sampling, allowing not only to obtain the temporal trace of the mean value of the THz field but also to characterize the behavior of its fluctuations.

The other two contributions $\Gamma_\mathrm{II}$ and $\Gamma_\mathrm{III}$ in Eq.~\eqref{result} constitute genuine quantum corrections that are partially missed by an effective theory for the nonlinear polarization. For the quantum corrections, the first interaction of the matter system with the THz field is given by the superoperator $\hat{E}_{\mathrm{THz},s,-}$, causing a perturbation in the THz quantum state which propagates further and interacts with a second molecule according to $\hat{E}_{\mathrm{THz},s,+}$. This process leads to an intermolecular time ordering of the two $\chi^{(2)}$ processes and an effective intermolecular interaction which is only understandable in the two-molecule space. This intermolecular time ordering is reflected in interference terms that can also result in a reduction of the electro-optic signal fluctuations. Contributions of this type involve the commutator of the THz field operators and are independent of the state of the THz field. They therefore represent byproducts of the nonlinear interaction that do not provide any information on the THz field.

$\Gamma_\mathrm{II}$ is given by
\begin{align}
\begin{split}
\Gamma_\mathrm{II}&=\left(\frac{N\omega_\mathsf{p}L}{c_0 }\right)^2\int_0^\infty \!\!\!d\Omega\frac{\hbar\Omega}{C}\Re\left\{D(\Omega,\theta)D_\mathrm{q}(\Omega,\theta)\right\}-2\frac{\hbar c_0}{CL}\Im\left\{D(\Omega,\theta)D_\mathrm{q}(\Omega,\theta)\right\},\label{gammaII}
\end{split}
\end{align}
where the classical detection window $D(\Omega,\theta)$ is given in Eq.~\eqref{class} and the detection window $D_\mathrm{q}(\Omega,\theta)$, which depends on quantum susceptibilities, is given by
\begin{align}
\begin{split}
D_\mathrm{q}(\Omega,\theta)&=\int_0^\infty\mathrm{d}\omega f^*_+(\omega,\Omega,\theta)\left(\chi^{(2)*}_{++-}(-\Omega;-\omega,\omega+\Omega)+\chi^{(2)*}_{+-+}(-\Omega;\omega+\Omega,-\omega)\right)\\
&\quad+\int_0^\infty\mathrm{d}\omega f^*_+(\omega,\Omega,\theta)\left(\chi^{(2)*}_{++-}(-\omega;-\Omega,\omega+\Omega)+\chi^{(2)*}_{+-+}(-\omega;\omega+\Omega,-\Omega)\right)\\
&\quad + \int_0^\infty\mathrm{d}\omega f_-(\omega,\Omega,\theta)\left(\chi^{(2)}_{++-}(\Omega;-\omega,\omega-\Omega)+\chi^{(2)}_{+-+}(\Omega;\omega-\Omega,-\omega)\right)\\
&\quad +\int_0^\infty\mathrm{d}\omega f_-(\omega,\Omega,\theta)\left(\chi^{(2)}_{++-}(-\omega;\Omega,\omega-\Omega)+\chi^{(2)}_{+-+}(-\omega;\omega-\Omega,\Omega)\right).\label{quant}
\end{split}
\end{align}
$D_\mathrm{q}$ involves quantum susceptibilities of the form $\chi^{(2)}_{++-}$ and $\chi^{(2)}_{+-+}$ [cf. Eq.\eqref{susc}]. The corresponding leading diagrams are depicted in Fig.~\ref{Fig2}($\mathrm{II}$a)-($\mathrm{II}$c). Observing features related to these susceptibilities therefore assures that a genuinely nonclassical field was involved in the responsible nonlinear process.

Finally, $\Gamma_\mathrm{III}$ is given by
\begin{align}
\begin{split}
\Gamma_\mathrm{III}&=\left(\frac{N\omega_\mathsf{p}L}{c_0 }\right)^2\int_0^\infty \!\!\!d\Omega\frac{\hbar\Omega}{C}\Re\left\{D(\Omega,\theta)D_\mathrm{casc}(\Omega,\theta)\right\}-2\frac{\hbar c_0}{CL}\Im\left\{D(\Omega,\theta)D_\mathrm{casc}(\Omega,\theta)\right\},\label{gammaIII}
\end{split}
\end{align}
where the detection window $D_\mathrm{casc}(\Omega,\theta)$, which depends on the classical susceptibilities $\chi^{(2)}_{+--}$, is given by
\begin{align}
\begin{split}
D_\mathrm{casc}(\Omega,\theta)&=\frac{1}{2}\int_0^\infty\mathrm{d}\omega f^*_+(\omega,\Omega,\theta)\left(\chi^{(2)*}_{+--}(-\Omega;-\omega,\omega+\Omega)+\chi^{(2)*}_{+--}(-\Omega;\omega+\Omega,-\omega)\right)\\
&\quad +\frac{1}{2}\int_0^\infty\mathrm{d}\omega f_-(\omega,\Omega,\theta)\left(\chi^{(2)}_{+--}(\Omega;-\omega,\omega-\Omega)+\chi^{(2)}_{+--}(\Omega;\omega-\Omega,-\omega)\right).\label{casc}
\end{split}
\end{align}
$\Gamma_\mathrm{III}$ constitutes a quantum correction even though the detection window $D_\mathrm{casc}(\Omega,\theta)$ only involves classical susceptibilities of the type $\chi^{(2)}_{+--}$. Here, we deal with cascading processes wherein for example a THz photon is emitted into the THz vacuum by molecule $b$ and is then reabsorbed at molecule $a$. The leading diagrams are shown in Figs.~\ref{Fig2}($\mathrm{III}$a) and ($\mathrm{III}$b). The cascading process can also be obtained within a macroscopic theory where it is treated phenomenologically using the Maxwell equations\cite{Mukamel1995}. However, the macroscopic treatment can only retrieve the first term of Eq.~\eqref{gammaIII}. The second term is related to the fact that the microscopic approach only enforces energy conservation for the entire cascaded process\cite{Bennett2014}. Instead, the macroscopic approach implicitly assumes that the spatial separation between the two involved $\chi^{(2)}$ processes is large enough so that energy is conserved for each single $\chi^{(2)}$ process.

\begin{figure}[ht!]
\includegraphics[width=\textwidth]{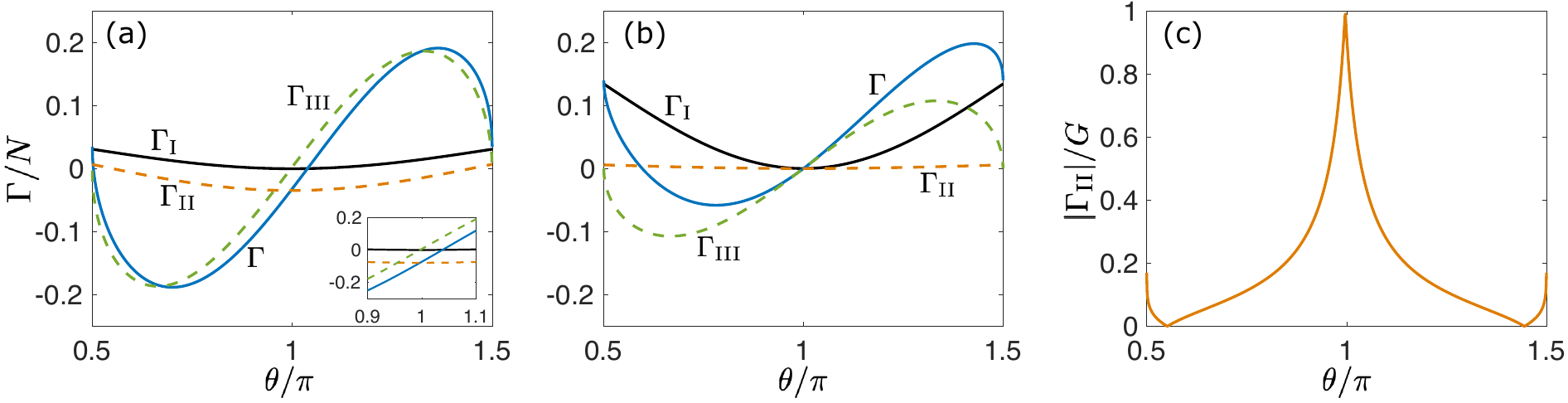}
\caption{Normally-ordered second moment $\Gamma$ of the electro-optic signal [calculated according to Eqs.~\eqref{gammaI}, \eqref{gammaII} and \eqref{gammaIII}] evaluated for the vacuum state of the THz field and for different phase shifts $\theta$. (a), case of the level scheme shown in Fig.~\ref{Fig1}(b). The blue line depicts $\Gamma$ for the full quantum treatment according to Eq.~\eqref{result}. It is given by the sum of the classical contribution $\Gamma_\mathrm{I}$ [black line, cf. Eq.~\eqref{gammaI}], the  contribution stemming from the quantum susceptibilities $\Gamma_\mathrm{II}$ [dashed orange line, cf. Eqs.~\eqref{gammaII}] and the cascading contribution $\Gamma_\mathrm{III}$ [dashed green line, cf. Eqs.~\eqref{gammaIII}]. The inset shows a zoom-in on the interval $0.9\pi\leq\theta\leq 1.1\pi$. (b), off-resonant case, where $\omega,\Omega\ll \omega'_{g'g},\omega'_{fg}$. The length $L$ of the nonlinear medium with respect to the strength of the nonlinear susceptibilities is chosen such that the maximal value of $\Gamma$ can still be regarded as a small correction to the shot noise (in this case a maximum contribution of $0.2N$ was adopted). (c), ratio of the absolute strength of the quantum contribution $|\Gamma_\mathrm{II}|$ to the combined absolute strengths of each contribution $G=|\Gamma_\mathrm{I}|+|\Gamma_\mathrm{II}|+|\Gamma_\mathrm{III}|$ for the level scheme shown in Fig.~\ref{Fig1}(b). The normally-ordered second moment $\Gamma$ is almost entirely determined by the quantum contribution $\Gamma_\mathrm{II}$ at $\theta$ being close to the half-wave plate configuration, $\theta=\pi$.\label{Fig3}}
\end{figure}

\section{SIMULATION RESULTS FOR THE THREE-LEVEL MODEL SCHEME}

To illustrate the difference between the classical and quantum response of matter, we have calculated them for the three-level model scheme shown in Fig.~\ref{Fig1}(b) and a sample length $L$ of $10$~ $\mu$m. We assume a rectangular spectral envelope of the probe field with a central frequency of $\omega_c/(2\pi)=255$~THz and a spectral width of $\Delta\omega/(2\pi)=150$~THz. Therefore, the probe is off-resonant with respect to both transition frequencies while the THz vacuum can include frequencies being resonant with the $g\rightarrow g'$ transition. Fig.~\ref{Fig3}(a) depicts the normally ordered second moment $\Gamma$ in Eq.~\eqref{result} together with its three components: the classical response $\Gamma_\mathrm{I}$, the response described by the quantum susceptibilities $\Gamma_\mathrm{II}$ and the response according to the cascading processes $\Gamma_\mathrm{III}$. $\Gamma_\mathrm{I}$ samples the THz vacuum fluctuations which are uncorrelated with the shot noise of the probe. Therefore, it can only enhance the fluctuations of the electro-optic signal. In contrast, both quantum corrections can also lead to a reduction of the noise due to the effective interaction between the molecules mediated by the THz vacuum. Interestingly, there are certain phase shifts $\theta$ whereby one of the three contributions is dominant. Figure~\ref{Fig3}(a) and (c) show that for a phase shift of $\theta=\pi/2$, corresponding to the experimental configuration with a quarter-wave plate, the quantum contributions are small compared to the classical response while for $\theta=\pi$, corresponding to the option with a half-wave plate, the quantum contribution $\Gamma_\mathrm{II}$ constitutes almost the entire correction to the unperturbed second moment of the electro-optic signal. Outcomes for intermediate phase shifts are dominated by the cascading contribution $\Gamma_\mathrm{III}$. The fact that both the classical and the cascading contributions almost vanish in the half-wave plate configuration ($\theta=\pi$) is remarkable and suggests that electro-optic sampling could also be used as a spectroscopic tool to study the pure quantum susceptibilities of different materials.

\begin{figure}[ht!]
\includegraphics[scale=1]{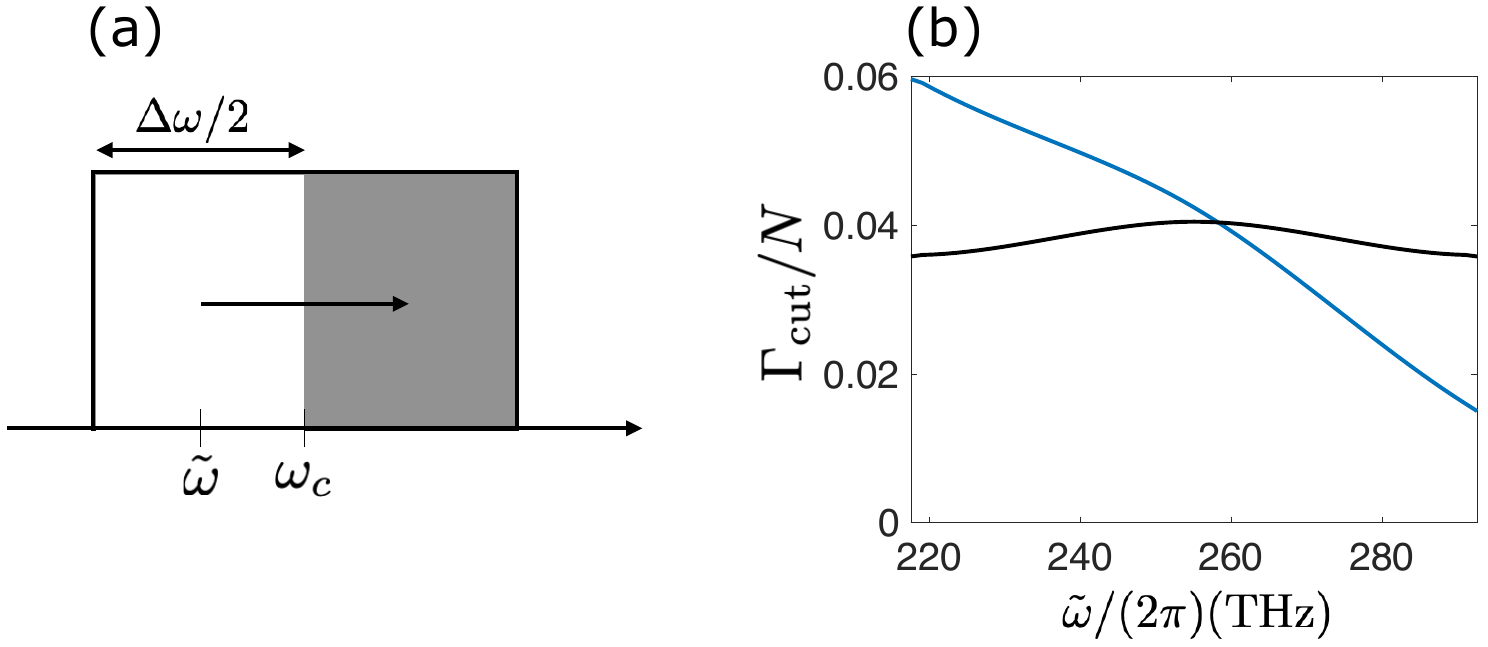}
\caption{Spectral filtering in electro-optic sampling and effects of the cascading processes. (a), the spectral filtering introduced in the detection window. For $\tilde{\omega}/(2\pi)=217.5$~THz, the lower half of the probe spectrum is being measured while  $\tilde{\omega}/(2\pi)=292.5$~THz constitutes a measurement of the upper half of the probe spectrum. (b), comparison of the normally-ordered second moment $\Gamma$ according to Eq.~\eqref{signalcut} between the classical treatment (black line), where only $\Gamma_\mathrm{I}$ is taken into account, and the full quantum treatment (blue line) involving also the cascading processes. \label{Fig4}}
\end{figure}

As demonstrated in Fig.~\ref{Fig3}(b), the quantum susceptibilities involved in the detection window $D_\mathrm{q}$ almost vanish in the off-resonant case. This might suggest that the classical approach usually serves as a good approximation when treating nonlinear processes involving quantum fields. The quantum correction, however, also involves cascading processes. The corresponding response is formed by the detection window $D_\mathrm{casc}$ which contains classical susceptibilities, too. Accordingly, the quantum correction to the classical response does not vanish completely, even if all involved fields are off-resonant with respect to the matter transition frequencies. Figure~\ref{Fig3}(b) shows the off-resonant normally-ordered second moment $\Gamma$ in Eq.~\eqref{result} where we have now assumed a three-level model with transition frequencies $\omega,\Omega\ll \omega'_{g'g},\omega'_{fg}$. The cascading processes can lead to substantial corrections for intermediate phase shifts of $\pi/2< \theta <\pi$ and $\pi< \theta < 3\pi/2$. These corrections are given by the second term in Eq.~\eqref{gammaIII}, which may not be derived macroscopically via Maxwell's equations. In previous experiments\cite{Riek2015,Benea2019}, the electro-optic sampling was executed for a phase shift of $\theta=\pi/2$. Interestingly, Fig.~\ref{Fig3}(b) shows that while the cascading processes result in distinct contributions for various phase shifts, they can be neglected and thus remain hidden in the quarter-wave plate configuration ($\theta=\pi/2$) so that the classical treatment provides a good approximation. However, their presence can be inferred, e.g. if spectral filtering is introduced into the measurement according to Fig.\ref{Fig4}(a) where only half of the probe spectrum is exploited for detection. The corresponding electro-optic signal is given by
\begin{align}
\hat{\mathcal{S}}_\mathrm{cut}(\tilde{\omega})=C\int_{\tilde{\omega}-\Delta\omega/4}^{\tilde{\omega}+\Delta\omega/4}\!\!\! \mathrm{d}\omega\; \frac{1}{\hbar\omega}\left[i\hat{E}^\dagger_{\mathrm{p},z}(\omega)\hat{E}_{\mathrm{p},s}(\omega)+H.c.\right],\label{signalcut}
\end{align}
where the frequency window is centered at $\tilde{\omega}$. Figure~\ref{Fig4}(a) depicts the spectral cut for $\tilde{\omega}=\omega_c-\Delta\omega/4$, which generally begins in the lower half of the probe spectrum at $\tilde{\omega}-\Delta\omega/4$ and ends in the upper half at $\tilde{\omega}+\Delta\omega/4$. Figure~\ref{Fig4}(b) illustrates how the normally-ordered second moment $\Gamma$ based on Eq.~\eqref{signalcut} behaves upon varying the position of $\tilde{\omega}$. To gain an additional insight, we now compare $\Gamma$ to the case when the cascading processes are excluded from the calculation. Note that the cascading processes add noise in the lower half of the probe spectrum while subtracting noise in the upper half of the probe spectrum. The corresponding contributions are cancelled out almost completely when the entire probe spectrum integrated over $\omega$ is detected.

\begin{figure}[ht!]
\includegraphics[scale=1]{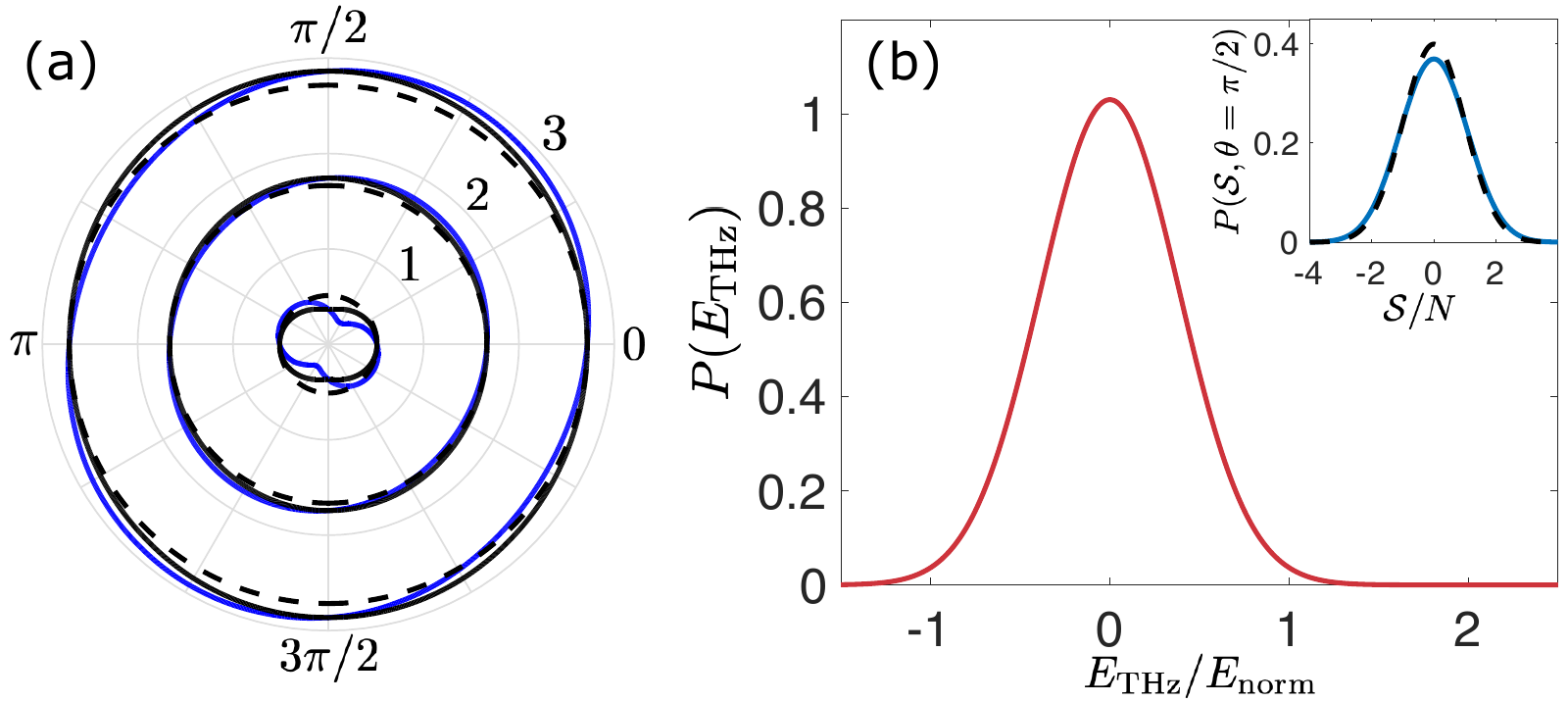}
\caption{Statistics of the measured electro-optic signal $\hat{\mathcal{S}}(\theta)$ [cf. Eq.~\eqref{signal}] and reconstructed statistics of the THz vacuum. (a), contour plot of the probability distribution of the electro-optic signal $\hat{\mathcal{S}}(\theta(\varphi))$. The blue line depicts the probability distribution according to Eq.~\eqref{result}, the black line shows the probability distribution resulting from the classical treatment, and the black dashed line depicts the probability distribution of the shot noise of the probe.  (b), reconstructed statistics of the THz vacuum with $E_\mathrm{norm}=\frac{c_0}{L\omega_\mathrm{p}\chi^{(2)}_{+--}}$ (the nonlinearity is independent of the frequency entries in the off-resonant case and can be treated as a constant prefactor). The blue line in the inset depicts the probability distribution of $\hat{\mathcal{S}}(\theta=\pi/2)$ and the dashed black line corresponds to the shot noise of the probe. Deconvolution of the two should provide the statistics of the THz vacuum. \label{Fig5}}
\end{figure}

We next use the results for $\Gamma$ to calculate the first correction due to the THz field to the probability distribution in Eq.~\eqref{prob}, which is given by
\begin{align}
\begin{split}
P(\mathcal{S},\theta)&=\frac{1}{\sqrt{2\pi N}}\exp\left(-\frac{\mathcal{S}^2}{2N}\right)\left(1+\frac{1}{2N^2}\left(\mathcal{S}^2-N\right)\Gamma\right).\label{probgamma}
\end{split}
\end{align}
The correction to the bare probability distribution of the shot noise of the probe is given by a single-photon state of the $\vec{e}_s$-polarized NIR field scaling linearly with the normally-ordered second moment of the electro-optic signal $\Gamma$ in Eq.~\eqref{result}. Figure~\ref{Fig5}(a) compares the contour plot of the probability distribution in Eq.~\eqref{probgamma} to that of the bare shot noise of the probe and the probability distribution obtained by only considering the classical contribution $\Gamma_\mathrm{I}$. Here, the angle in the polar coordinates does not represent $\theta$, which corresponds to the phase shift induced by the waveplate shown in Fig.~\ref{Fig1}(a), but rather the phase shift $\varphi=\arccos\left(\sqrt{-\cos\theta}\right)$ induced in the electro-optic signal itself, given by $e^{i\varphi(\theta)}=P(\theta)$ [cf. Eq.~\eqref{signal}]. The noise added by the classical contribution $\Gamma_\mathrm{I}$ turns the rotationally symmetric probability distribution of the shot noise of the probe into an ellipse with the long axis along the $\varphi=\pi/2$ direction. The quantum correction given mainly by the cascading processes $\Gamma_\mathrm{III}$ then rotates the long axis of this ellipse and squeezes the generated NIR field $\hat{E}_{\mathrm{p},s}$ for phase shifts of $0.7\pi\lesssim\varphi\leq\pi$.

Ellipsometry enables a full tomography of the generated NIR field $\hat{E}_{\mathrm{p},s}$, as in standard balanced homodyne detection\cite{schleich_book}. We now discuss whether the statistics of the THz field itself can be reconstructed from the statistics of the generated NIR field. Figure~\ref{Fig5}(b) shows this type of reconstruction for a phase shift of $\varphi=\pi/2$. In this case, the quantum corrections to $\Gamma$ are negligible and the noise added on top of the shot noise is mainly given by the classical contribution $\Gamma_\mathrm{I}$ which samples the fluctuations of the THz field and is therefore uncorrelated with the shot noise. Here, the probability distribution of the THz field is given by a simple deconvolution of the probability distribution in Eq.~\eqref{probgamma} with the Gaussian probability distribution of the shot noise of the probe. In contrast, the quantum correction dominated by the cascading processes $\Gamma_\mathrm{III}$ introduces an additional contribution for other phase shifts that is independent on the state of the THz field, i.e. it cannot be understood as simply sampling the THz vacuum fluctuations. The cascading processes lead to interference with the $\vec{e}_s$-polarized NIR field $\hat{E}_{\mathrm{p},s}$ and therefore with the shot noise. A simple deconvolution of the obtained probability distribution for the corresponding phase shifts $\varphi$ would fail to give the correct statistics of the THz vacuum.

\section{CONCLUSIONS}
We have employed the superoperator formalism to develop a microscopic theory of time-domain electro-optic sampling of quantum fluctuations of electric fields. We identified three contributions to the electro-optic signal variance, given by a classical contribution reproducing the results of previous theoretical models\citep{Moskalenko2015} and two quantum responses: a contribution stemming from quantum susceptibilities and a contribution describing cascaded nonlinear processes. The quantum responses strongly depend on intermolecular time-ordering of the THz interactions in the $\chi^{(2)}$ processes, where an effective interaction between the molecules, mediated by the THz vacuum fluctuations, is established. To demonstrate the difference between the classical and quantum response, we compared the respective phase-dependent electro-optic contributions for an effective nonlinear medium described by a system of 3-level non-interacting molecules. For a configuration involving a quarter-wave plate, we found that the quantum contributions to the normally-ordered second moment of the electro-optic signal vanish. Here, an effective theory for the nonlinear interaction which relies solely on the classical response of matter constitutes a good approximation. For other phase shifts, however, the quantum response significantly influences the measured fluctuations and can even lead to a substantial reduction of the fluctuations below the shot-noise limit of the probe field. For the configuration with a half-wave plate, the classical as well as the cascading processes can be suppressed completely, which opens up the possibility to use electro-optic sampling as a novel spectroscopic tool to study quantum susceptibilities only accessible through truly non-classical states of light. Because of the cascading processes involved in the quantum response, traces of the quantum correction should already be observable with slight changes in present experimental setups by either introducing spectral filtering or studying the electro-optic signal for certain phase shifts in the ellipsometry part. Finally, we describe how the probability distribution of the THz vacuum can be reconstructed from the probability distribution for the electro-optic signal as has been achieved experimentally\cite{Riek2015}. We note that a reconstruction for other phase shifts would need to take into account the additional quantum contributions. We have shown that the ellipsometry enables the detection of the probability distribution of the generated field for phase shifts within an interval of length $\pi$, thus enabling a full quantum tomography of its state. A method for reconstructing the statistics of the sampled THz field for arbitrary phase shifts would pave a way for quantum tomography of nonclassical fields with subcycle temporal resolution. Our theory provides a firm basis for addressing the challenging issue of the  extraction of these statistics under realistic experimental conditions.

\section*{Acknowledgements}
 S. M. gratefully acknowledges the support of the National Science Foundation Grant CHE-1953045. M.K. is indebted to the LGFG PhD fellowship programme of the University of Konstanz. M.K. and A.S.M. were supported by the Baden-Württemberg Stiftung via the Elite Programme for Postdocs. A.S.M. was also supported by the National Research Foundation of Korea (NRF) grant funded by the Korea government (MSIT) (2020R1A2C1008500).

\newpage

\renewcommand{\theequation}{A\arabic{equation}}
\renewcommand{\thefigure}{A\arabic{figure}}
\setcounter{equation}{0}
\setcounter{figure}{0}
\setcounter{section}{0}

\setcounter{secnumdepth}{3}
\renewcommand{\thesection}{APPENDIX \Alph{section}}

\section{ELLIPSOMETRY AND DERIVATION OF THE SIGNAL PROBABILITY DISTRIBUTION\label{AppA}}
Here we derive Eq.~\eqref{signal} and describe the setup required to characterize the electro-optic signal for arbitrary phase shifts of the waveplate in Fig.~\ref{Fig1}. We consider a tunable waveplate that can introduce arbitrary phase shifts $\theta$. If the fast axis of the waveplate is rotated by an angle $\alpha$ against $\vec{e}_z$, the total NIR electric field after the waveplate can be expressed as\cite{Sulzer2020}
\begin{align}
\left(
\begin{array}{c}
\hat{E}'_{\mathrm{p},s}\\
\hat{E}'_{\mathrm{p},z}\\
\end{array}
\right)=\left(
\begin{array}{c c}
\cos\alpha&\sin\alpha\\
-\sin\alpha&\cos\alpha\\
\end{array}
\right)\left(
\begin{array}{c c}
e^{-i\theta/2}&0\\
0&e^{i\theta/2}\\
\end{array}
\right)\left(
\begin{array}{c c}
\cos\alpha&-\sin\alpha\\
\sin\alpha&\cos\alpha\\
\end{array}
\right)\left(
\begin{array}{c}
\hat{E}_{\mathrm{p},s}\\
\hat{E}_{\mathrm{p},z}\\
\end{array}
\right),
\end{align}
where $\hat{E}_{\mathrm{p},s},\hat{E}_{\mathrm{p},z}$ are the polarization components of the probe field arriving at the waveplate whereas $\hat{E}'_{\mathrm{p},s},\hat{E}'_{\mathrm{p},z}$ represent these components at the output of the waveplate.

After the waveplate, the photon numbers in the $z$ and $s$ polarization are measured. The electro-optic signal is given by the difference
\begin{align}
\hat{\mathcal{S}}(\theta)&\equiv\hat{N}'_z-\hat{N}'_s=C\int_0^\infty\mathrm{d}\omega \frac{1}{\hbar\omega}\left(\hat{E}'^\dagger_{\mathrm{p},z}(\omega)\hat{E}'_{\mathrm{p},z}(\omega)-\hat{E}'^\dagger_{\mathrm{p},s}(\omega)\hat{E}'_{\mathrm{p},s}(\omega)\right)\\
&= C\int_0^\infty\mathrm{d}\omega \frac{1}{\hbar\omega}\left[\left(\sin^2\frac{\theta}{2}\sin 4\alpha+i\sin\theta\sin 2\alpha\right)\hat{E}^\dagger_{\mathrm{p},z}(\omega)\hat{E}_{\mathrm{p},s}(\omega)+H.c.\right]\\
&\quad+\left(\cos^2\frac{\theta}{2}+\sin^2\frac{\theta}{2}\cos 4\alpha\right)\left(\hat{N}_z-\hat{N}_s\right),\label{dersig}
\end{align}
where the operator $\hat{N}_z-\hat{N}_s$ denotes the difference in photon numbers before the waveplate. For a balanced detection, the second term on the right hand side of Eq.~\eqref{dersig} must vanish. This can be achieved by adjusting the angle of the fast axis $\alpha$ for each phase shift $\theta$ so that
\begin{align*}
\cos^2\frac{\theta}{2}+\sin^2\frac{\theta}{2}\cos 4\alpha=0.
\end{align*}
This condition is satisfied by $\alpha=\arccos\left(-\cot^2\frac{\theta}{2}\right)/4$ for $\pi/2\leq\theta\leq3\pi/2$. The balanced electro-optic signal is then given by
\begin{align*}
\hat{\mathcal{S}}(\theta)=C\int_0^\infty\!\!\! \mathrm{d}\omega\; \frac{1}{\hbar\omega}\left[P(\theta)\hat{E}^\dagger_{\mathrm{p},z}(\omega)\hat{E}_{\mathrm{p},s}(\omega)+H.c.\right],
\end{align*}
with $P(\theta)=\sqrt{-\cos(\theta)}+i\sqrt{1+\cos(\theta)}$.\\
We next derive the probability distribution $P(\mathcal{S},\theta)$ to measure a certain value $\mathcal{S}$ for the electro-optic signal $\hat{\mathcal{S}}(\theta)$ for a given phase shift $\theta$. We follow the same steps as Raymer \textit{et al.}\cite{Raymer1995}. The electro-optic signal is obtained by measuring the difference between the detected numbers of the $\vec{e}_z$-polarized and $\vec{e}_s$-polarized photons. We therefore start with the joint probability to measure $n_z$ $\vec{e}_z$-polarited-photons and $n_s$ $\vec{e}_s$-polarized photons,
\begin{align}
\begin{split}
P(n_z,n_s)&=\left\langle :\! \frac{\hat{N}_z^{\prime n_z}}{n_z!}e^{-\hat{N}^\prime_z}\frac{\hat{N}_s^{\prime n_s}}{n_s!}e^{-\hat{N}^\prime_s}\!:\right\rangle\\
&\approx\left\langle :\! \frac{1}{\sqrt{2\pi\hat{N}^\prime_z}}e^{-\frac{\left(n_z-\hat{N}_z^\prime\right)^2}{2\hat{N}_z^\prime}}\frac{1}{\sqrt{2\pi\hat{N}^\prime_s}}e^{-\frac{\left(n_s-\hat{N}_s^\prime\right)^2}{2\hat{N}_s^\prime}}\!:\right\rangle,
\end{split}
\end{align}
where we assumed a lossless detector and approximated the Poissonian function as a Gaussian in the second line because of the strong coherent probe resulting in large expected photon numbers for both polarization directions. We are interested in the probability to measure $\mathcal{S}=n_z-n_s$, which can be derived according to
\begin{align*}
P(\mathcal{S})&=\sum_{n_z} P(n_z,n_z-\mathcal{S})\\
&=\int\mathrm{d}x P(x,x-\mathcal{S})\\
&=\left\langle :\! \frac{1}{\sqrt{2\pi\hat{N}^\prime_\mathrm{T}}}e^{-\frac{\left(\mathcal{S}-\hat{\mathcal{S}}(\theta)\right)^2}{2\hat{N}^\prime_\mathrm{T}}}\!:\right\rangle,
\end{align*}
with $\hat{N}^\prime_\mathrm{T}=\hat{N}^\prime_z+\hat{N}^\prime_s$ as the total detected photon number. Here, we replaced the discrete sum by an integral because of the large expected total photon number and then used the fact that a convolution of two Gaussians also results in a Gaussian distribution. The total detected photon  number can then be well approximated by the mean number of photons in the probe $N$, resulting in
\begin{align}
P(\mathcal{S},\theta)&=\left\langle :\! \frac{1}{\sqrt{2\pi N}}e^{-\frac{\left(\mathcal{S}-\hat{\mathcal{S}}(\theta)\right)^2}{2N}}\!:\right\rangle.
\end{align}
Expanding the exponential in orders of the electro-optic signal $\hat{\mathcal{S}}$ finally leads to Eq.~\eqref{prob}.

\renewcommand{\theequation}{B\arabic{equation}}
\renewcommand{\thefigure}{B\arabic{figure}}
\section{THE TOTAL ELECTRO-OPTIC SIGNAL; DETAILED CALCULATION OF THE DIAGRAMS\label{AppB}}
Here we give a detailed example for the calculation of the normally-ordered second moment of the signal $\Gamma$ for a general quantum THz field with a vanishing mean field, given by the density matrix $\hat{\rho}_\mathrm{THz}$. In this case, the initial field density matrix is given by $\hat{\rho}_\mathrm{field}=\Ket{\{\mathcal{E}_{\mathrm{p},z}\}}\!\Bra{\{\mathcal{E}_{\mathrm{p},z}\}}\otimes \Ket{0_{\mathrm{NIR},s}}\!\Bra{0_{\mathrm{NIR},s}}\otimes\hat{\rho}_\mathrm{THz}$. In order to obtain a non-vanishing contribution to the normally-ordered second moment $\Gamma$ in Eq.~\eqref{Gamma}, we need to expand the exponential to sixth order. The expectation value over the matter degrees of freedom may then be split into the product of two expectation values for two different molecules where each molecule interacts once with each of the three fields.

\begin{figure}[!htbp]
\centering
\includegraphics[width=\textwidth]{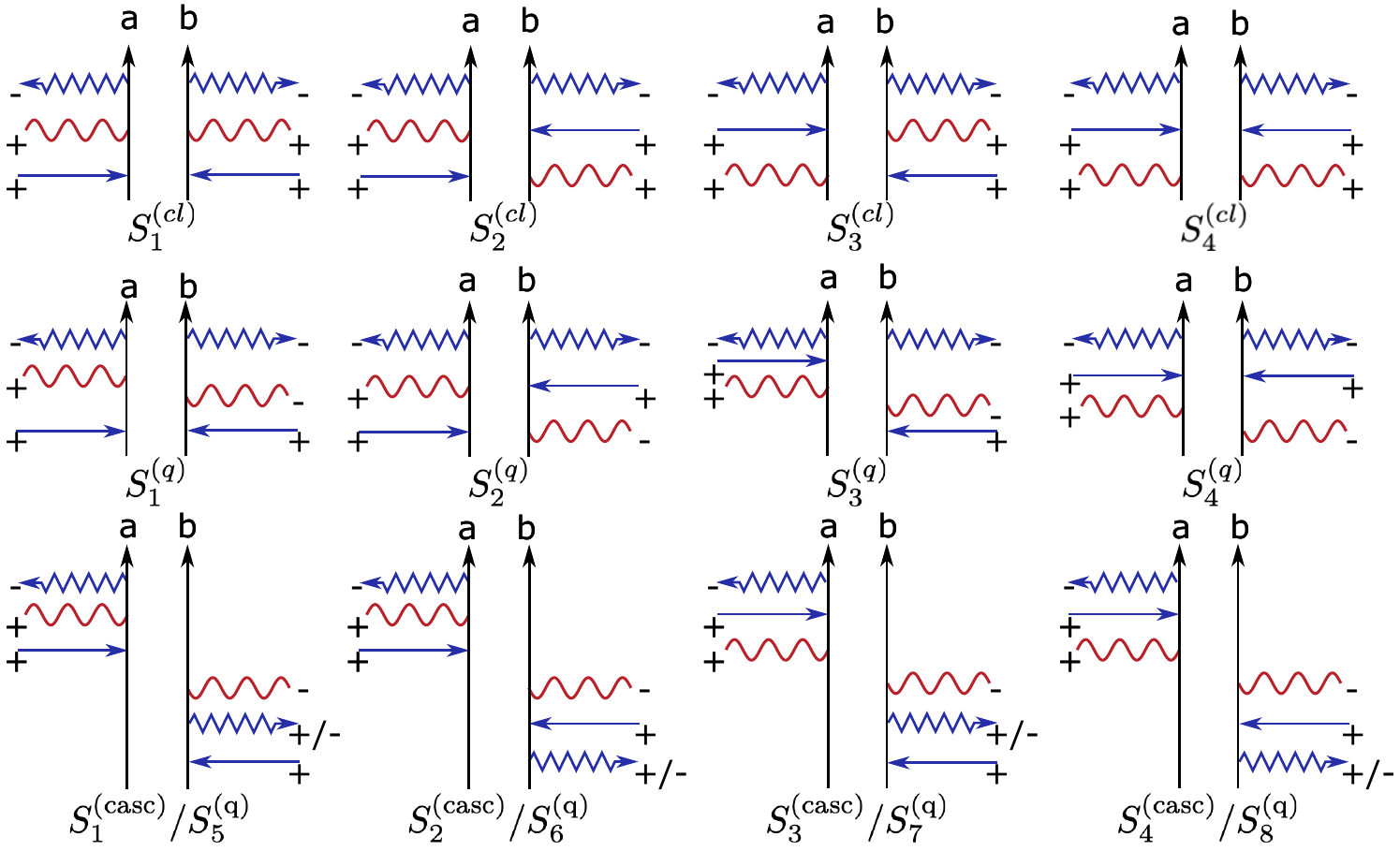}
\caption{Complete set of superoperator diagrams contributing to $\Gamma$[cf. Eq.~\eqref{Gamma}. Vertical arrows represent the density matrices of a pair molecules $a$ and $b$ as indicated, with the past at the bottom and the future at the top. Horizontal arrows denote interactions with the electric fields. Blue zigzag arrows: generated NIR field $\hat{E}_{\mathrm{p},s}$; wavy red lines: THz field $\hat{E}_{\mathrm{THz},s}$; straight blue arrows: coherent probe field $\hat{E}_{\mathrm{p},z}$. An arrow pointing to the right (left) corresponds to the annihilation (creation) of a photon. The interaction with the THz field (wavy red lines) does not have an arrow since for any interaction it can be pointing either to the right or to the left. The $\pm$ indices denote the superoperator nature of the corresponding mode of the interacting field. Additional contributing diagrams are given by flipping all field arrows attributed to molecule $a$ and/or molecule $b$. Note that flipping the field arrows at both molecules results in the Hermitian conjugation of the corresponding terms. The first row describes the classical DFG response contribution to $\Gamma$ if the THz arrows are chosen to point outwards at each molecule while the classical SFG is obtained when all of the THz arrows point inwards. Here, the intermolecular time ordering does not play any role. The second and third rows represent the quantum corrections to the classical response. Terms with the superscript $(\mathrm{q})$ denote processes involving quantum susceptibilities on molecule $b$ (note the two $'-'$-type interactions) while terms with the superscript $(\mathrm{casc})$ describe cascading processes.  For the second and third row, the intermolecular time ordering is important since the interaction with the THz field on molecule $a$ has to always occur after the THz interaction with molecule $b$.\label{S1}}
\end{figure}

To derive the nonlinear response of matter and pinpoint its quantum nature, we work with the $'\pm'$ superoperator algebra\cite{Mukamel1995}. Fig.~\ref{S1} depicts all of the diagrams contributing to $\Gamma$ which is given in Eq.~\eqref{Gamma}.
Since the probe field is classical because of its strong coherent amplitude, we can neglect any quantum features by only using the $'+'$-superoperator $\hat{E}_{\mathrm{p},z,+}$ for this field. This step means that it can be simply replaced by its coherent amplitude $\mathcal{E}_{\mathrm{p},z}$. The generated NIR  field $\hat{E}_{\mathrm{p},s}$ and the THz field $\hat{E}_{\mathrm{THz},s}$ are essentially quantum, i.e. they can appear with either superoperator index $'+'$ or $'-'$. The electro-optic signal in Eq.~\eqref{signal} describes the measurement of the generated NIR field. Consequently, the last interaction has to also occur with $\hat{E}_{\mathrm{p},s,-}$ for obtaining a non-vanishing expectation value for the matter degrees of freedom. Any other interaction would lead to a trace over a commutator for either the field or matter expectation value which vanishes. The previous interaction with the generated NIR probe field at the other molecule can then either be a $'+'$ or $'-'$. Contrary to the generated NIR probe field, the last interaction with the THz field must be a $'+'$ to obtain a non-vanishing expectation value for the field degrees of freedom, since this field is not being measured. Again the previous interaction with the other molecule can either be a $'+'$ or $'-'$. Finally, the last interaction at each molecule must be a $'-'$ for the field to obtain a non-vanishing expectation value for the matter degrees of freedom. These considerations leave us with the contributing diagrams shown in Fig.~\ref{S1}.

The first row depicts the classical matter response where the interaction of matter with the THz field is considered classical (because it is a $'+'$ for the field). In general, the response depends on the actual state of the THz field. The second row is essentially the same as the first with the difference that molecule $b$ interacts with the THz mode according to the $'-'$ type. In that case, the response of molecule $b$ may not be described by a classical susceptibility. Here, intermolecular time ordering is crucial since these diagrams only contribute to the electro-optic signal if the interaction with the THz field on molecule $b$ takes place before molecule $a$. The third row depicts diagrams where the last interaction with molecule $b$ is with the THz field. That field must then be considered quantum (i.e. has to have a $'-'$) in order to obtain a non-vanishing expectation value for the field degrees of freedom for molecule $b$. This finding means that the entire $\chi^{(2)}$ process on molecule $b$ has to take place before the interaction with the THz field on molecule $a$. The diagrams in the third row where the measured field interacts with molecule $b$ according to a $'+'$ ($\hat{E}_{\mathrm{p},s,+}$)  can be interpreted as a cascaded $\chi^{(2)}$ process\cite{Bennett2014} where e.g. a THz photon is emitted during the first $\chi^{(2)}_{+--}$ interaction and is afterwards absorbed in a second $\chi^{(2)}_{+--}$ process emitting a NIR photon back into the detected field $\hat{E}_{\mathrm{p},s}$. The contributions to $\Gamma$ given by the second and third row in Fig.~\ref{S1} involve the commutator of the THz field, essentially amounting to a factor proportional to its frequency $\Omega$ and are therefore independent on the actual state of the THz field.

To illustrate the difference between the classical and the quantum response, let us write down the contributions for diagrams $S^\mathrm{(cl)}_4$ and $S^\mathrm{(\mathrm{casc})}_4$ in Fig.~\ref{S1}. We assume that the nonlinear medium consists of noninteracting molecules at uniform density and has a length of $L$ along the propagation direction of the involved electric fields. The coherent amplitude of the probe field is represented by
\begin{align*}
\mathcal{E}_{\mathrm{p},z}(\vec{r},t)=\int_0^\infty\mathrm{d}\omega \mathcal{E}_{\mathrm{p},z}(\omega)e^{i(k_\omega x-\omega t)},
\end{align*}
and the quantized electric field is given by
\begin{align*}
\hat{E}(\vec{r},t)&=\int_0^\infty\mathrm{d}\omega \hat{E}(\omega)e^{i(k_\omega x-\omega t)}\\
&=-i\int_0^\infty\mathrm{d}\omega\sqrt{\frac{\hbar\omega}{C}}\hat{a}(\omega)e^{i(k_\omega x-\omega t)}.
\end{align*}
Here, $\hat{E}(\vec{r},t)$ corresponds to the annihilation operator in the time domain, which is usually expressed as the positive frequency field $\hat{E}^{(+)}(\vec{r},t)$ in standard quantum optics notation\cite{Glauber1963}. We choose not to follow this convention here in order to avoid confusion with the superoperator notation. All fields are considered as plane waves propagating along the $x$-axis.

Diagram $S^\mathrm{(cl)}_4$ for outwards-pointing THz field arrows, corresponding to two DFG processes, is given by
\begin{align*}
\begin{split}
S^\mathrm{(cl)}_4&=\frac{-1}{\hbar^6}\int_V\mathrm{d}\vec{r}_a\int_V\mathrm{d}\vec{r}_b\int_{-\infty}^\infty\mathrm{d}\tau_3\int_{-\infty}^{\tau_3}\mathrm{d}\tau_2\int_{-\infty}^{\tau_2}\mathrm{d}\tau_1\int_{-\infty}^\infty\mathrm{d}\overline{\tau}_3\int_{-\infty}^{\overline{\tau}_3}\mathrm{d}\overline{\tau}_2\int_{-\infty}^{\overline{\tau}_2}\mathrm{d}\overline{\tau}_1\\
&\times\mathrm{tr}\left\{:\!\hat{\mathcal{S}}^2(\theta)\!\!:\hat{E}^\dagger_{\mathrm{p},s,-}(\vec{r}_a,\tau_3)\hat{E}_{\mathrm{p},z,+}(\vec{r}_a,\tau_2)\hat{E}^\dagger_{\mathrm{THz},s,+}(\vec{r}_a,\tau_1)\hat{E}_{\mathrm{p},s,-}(\vec{r}_b,\overline{\tau}_3)\hat{E}^\dagger_{\mathrm{p},z,+}(\vec{r}_b,\overline{\tau}_2)\hat{E}_{\mathrm{THz},s,+}(\vec{r}_b,\overline{\tau}_1)\hat{\rho}_\mathrm{field}\right\}\\
& \times\mathrm{tr}\left\{\hat{V}_{s,+}(\tau_3)\hat{V}_{z,-}(\tau_2)\hat{V}_{s,-}(\tau_1)\hat{\rho}_{\mathrm{mat},a}\right\}\mathrm{tr}\left\{\hat{V}_{s,+}(\overline{\tau}_3)\hat{V}_{z,-}(\overline{\tau}_2)\hat{V}_{s,-}(\overline{\tau}_1)\hat{\rho}_{\mathrm{mat},b}\right\}
\end{split}\\
&=2\left(\frac{N\omega_\mathrm{p}}{2\pi\varepsilon_0 c_0\hbar^2}\right)^2\int_0^\infty \mathrm{d}\omega\mathrm{d}\omega'\mathrm{d}\omega_2\mathrm{d}\omega_1\mathrm{d}\Omega\mathrm{d}\Omega'~\mathbb{E}(\omega,\omega',\omega_2,\omega_1)\int_{-\infty}^\infty\mathrm{d}\tau_3\mathrm{d}\overline{\tau}_3\int_0^\infty\mathrm{d}t_2\mathrm{d}t_1\mathrm{d}\overline{t}_2\mathrm{d}\overline{t}_1\\
&\qquad \times e^{i(\omega+\Omega-\omega_1)\tau_3}e^{i(\omega_1-\Omega)t_2}e^{-i\Omega t_1}e^{-i(\omega'+\Omega'-\omega_2)\overline{\tau}_3}e^{-i(\omega_2-\Omega')\overline{t}_2}e^{i\Omega'\overline{t}_1}\Xi_{+--;+--}(t_2,t_1;\overline{t}_2,\overline{t}_1)\\
&\qquad\qquad \times\int_0^L\mathrm{d}x_a\int_0^L\mathrm{d}x_b e^{-i(k_\omega+k_\Omega-k_{\omega_1})x_a}e^{i(k_{\omega'}+k_{\Omega'}-k_{\omega_2})x_b}\mathrm{tr}\left\{\hat{E}^\dagger_{\mathrm{THz},s}(\Omega)\hat{E}_{\mathrm{THz},s,+}(\Omega')\hat{\rho
}_\mathrm{THz}\right\},
\end{align*}
where
\begin{align*}
\mathbb{E}(\omega,\omega',\omega_2,\omega_1)&=\frac{\mathcal{E}^*_{\mathrm{p},z}(\omega)\mathcal{E}_{\mathrm{p},z}(\omega')\mathcal{E}^*_{\mathrm{p},z}(\omega_2)\mathcal{E}_{\mathrm{p},z}(\omega_1)}{4\left(\int_0^\infty\mathrm{d}\omega\left|\mathcal{E}_{\mathrm{p},z}(\omega)\right|^2\right)^2}
\end{align*}
results from replacing the probe field operators $\hat{E}_{\mathrm{p},z}$ by their coherent amplitudes (thereby neglecting any quantum effects resulting from these fields), $\omega_\mathrm{p}=\int_0^\infty\mathrm{d}\omega\left|\mathcal{E}_{\mathrm{p},z}(\omega)\right|^2/\int_0^\infty\mathrm{d}\omega\omega^{-1}\left|\mathcal{E}_{\mathrm{p},z}(\omega)\right|^2$ is the average detected frequency and $N=C\int_0^\infty\mathrm{d}\omega\frac{1}{\hbar\omega}\left|\mathcal{E}_{\mathrm{p},z}(\omega)\right|^2$ is the mean photon number of the coherent probe, coinciding with its variance which represents the shot noise. The expression
\begin{align*}
\Xi_{+--;+--}(t_2,t_1;\overline{t}_2,\overline{t}_1)&=\mathrm{tr}\left\{\hat{V}_{s,+}\hat{G}(t_2)\hat{V}_{z,-}
\hat{G}(t_1)\hat{V}_{s,-}\hat{\rho}_{\mathrm{mat},a}\right\}\\
&\qquad\qquad\times\mathrm{tr}\left\{\hat{V}_{s,+}\hat{G}(\overline{t}_2)\hat{V}_{z,-}\hat{G}(\overline{t}_1)\hat{V}_{s,-}\hat{\rho}_{\mathrm{mat},b}\right\}
\end{align*}
is given by the traces over the density matrices of the two molecules $a$ and $b$ according to the 3-level model shown in Fig.~\ref{Fig1}(b). Here,  $\hat{G}(t)=\exp\left(\frac{-i}{\hbar}\int_{-\infty}^t\mathrm{d}t \hat{H}_{\mathrm{mat},-}(t)\right)$ with $\hat{H}_\mathrm{mat}=\sum_{i=g',f}\hbar\omega_{ig}\Ket{i}\!\Bra{i}$ is the Green function superoperator of the matter system. \\
Finally, $S^\mathrm{(cl)}_4$ is given by
\begin{align*}
S^\mathrm{(cl)}_4&=2\left(\frac{N\omega_\mathrm{p}L}{c_0}\right)^2\int_0^\infty \mathrm{d}\omega\mathrm{d}\omega'\mathrm{d}\Omega\mathbb{E}(\omega,\omega',\omega'+\Omega,\omega+\Omega)\mathrm{tr}\left\{\hat{E}^\dagger_{\mathrm{THz},s}(\Omega)\hat{E}_{\mathrm{THz},s,+}(\Omega')\hat{\rho
}_\mathrm{THz}\right\}\\
&\qquad\qquad\qquad\times\mathrm{tr}\left\{\hat{V}_{s,+}\hat{G}(\omega)\hat{V}_{z,-}\hat{G}(-\Omega)\hat{V}_{s,-}\hat{\rho}_{\mathrm{mat},a}\right\}\mathrm{tr}\left\{\hat{V}_{s,+}\hat{G}(\omega')\hat{V}_{z,-}\hat{G}(-\Omega)\hat{V}_{s,-}\hat{\rho}_{\mathrm{mat},b}\right\}^*
\end{align*}
To obtain the expression for inwards-pointing arrows (two SFG processes) we simply need to change the sign of $\Omega$ in every argument and use the general relation $\hat{E}_{\mathrm{THz},s}(-\Omega)=\hat{E}_{\mathrm{THz},s}^\dagger(\Omega)$.

For $S^\mathrm{(\mathrm{casc})}_4$, we calculate the interaction with $\hat{E}_{\mathrm{p},s,+}$ on molecule $b$ and sum up the diagrams where the THz arrows are pointing outwards and inwards.The result is
\begin{align*}
S^\mathrm{(\mathrm{casc})}_4&=\frac{-1}{\hbar^6}\int_V\mathrm{d}\vec{r}_a\int_V\mathrm{d}\vec{r}_b\int_{-\infty}^\infty\mathrm{d}\tau_6\int_{-\infty}^{\tau_6}\mathrm{d}\tau_5\int_{-\infty}^{\tau_5}\mathrm{d}\tau_4\int_{-\infty}^{\tau_4}\mathrm{d}\tau_3\int_{-\infty}^{\tau_3}\mathrm{d}\tau_2\int_{-\infty}^{\tau_2}\mathrm{d}\tau_1\\
&\qquad\times\mathrm{tr}\left\{:\!\hat{\mathcal{S}}^2(\theta)\!\!:\hat{E}^\dagger_{\mathrm{p},s,-}(\vec{r}_a,\tau_6)\hat{E}_{\mathrm{p},z,+}(\vec{r}_a,\tau_5)\hat{E}^\dagger_{\mathrm{p},z,+}(\vec{r}_b,\tau_2)\hat{E}_{\mathrm{p},s,+}(\vec{r}_b,\tau_1)\right.\\
&\qquad\qquad\quad\times\left.\left[\hat{E}_{\mathrm{THz},s,+}(\vec{r}_a,\tau_4)\hat{E}^\dagger_{\mathrm{THz},s,-}(\vec{r}_b,\tau_3)+\hat{E}^\dagger_{\mathrm{THz},s,+}(\vec{r}_a,\tau_4)\hat{E}_{\mathrm{THz},s,-}(\vec{r}_b,\tau_3)\right]\hat{\rho}_\mathrm{field}\right\}\\
&\qquad \times\mathrm{tr}\left\{\hat{V}_{s,+}(\tau_6)\hat{V}_{z,-}(\tau_5)\hat{V}_{s,-}(\tau_4)\hat{\rho}_{\mathrm{mat},a}\right\}\mathrm{tr}\left\{\hat{V}_{s,+}(\tau_3)\hat{V}_{z,-}(\tau_2)\hat{V}_{s,-}(\tau_1)\hat{\rho}_{\mathrm{mat},b}\right\}\\
&=-\left(\frac{N\omega_\mathrm{p}}{2\pi\varepsilon_0 c_0\hbar^2}\right)^2\int_0^\infty \mathrm{d}\omega\mathrm{d}\omega'\mathrm{d}\omega_2\mathrm{d}\omega_1\mathrm{d}\Omega~\mathbb{E}(\omega,\omega',\omega_2,\omega_1)\frac{\hbar\Omega}{C}\int_{-\infty}^\infty\mathrm{d}\tau_6\int_0^\infty\mathrm{d}t_5\mathrm{d}t_4\mathrm{d}t_3\mathrm{d}t_2\mathrm{d}t_1\\
&\qquad \times e^{i(\omega-\omega_1-\omega'+\omega_2)\tau_6}e^{i(\omega_1+\omega'-\omega_2)t_5}e^{i(\omega'-\omega_2)t_4}e^{i(\omega'-\omega_2)t_2}e^{i\omega' t_1}\Xi_{+--;+--}(t_5,t_4;t_2,t_1)\\
&\qquad\times\int_0^L\mathrm{d}x_a\int_0^L\mathrm{d}x_b e^{-i(k_\omega-k_{\omega_1})x_a}e^{i(k_\omega'-k_{\omega_2})x_b}\left[e^{i(k_\Omega(x_a-x_b)-\Omega t_3)}-e^{-i(k_\Omega(x_a-x_b)-\Omega t_3)}\right]e^{i(\omega'-\omega_2)t_3}.
\end{align*}
Note that $S^\mathrm{(\mathrm{casc})}_4$ involves a completely time-ordered expression where the full $\chi^{(2)}$ process on molecule $b$ needs to occur before the $\chi^{(2)}$ process on molecule $a$ starts. In comparison, the two $\chi^{(2)}$ processes in $S^\mathrm{(cl)}_4$ happen completely independent of each other, i.e. there is no intermolecular time ordering. Finally, $S^\mathrm{(\mathrm{casc})}_4$ is given by
\begin{align*}
S^\mathrm{(\mathrm{casc})}_4&=-\left(\frac{N\omega_\mathrm{p}L}{c_0}\right)^2\int_0^\infty \mathrm{d}\omega\mathrm{d}\omega'\mathrm{d}\overline{\omega}\mathbb{E}(\omega,\omega',\overline{\omega},\omega+\overline{\omega}-\omega')\left(\frac{\hbar(\omega'-\overline{\omega})}{2C}-i\frac{\hbar c_0}{LC}\right)\\
&\qquad\times\mathrm{tr}\left\{\hat{V}_{s,+}\hat{G}(\omega)\hat{V}_{z,-}\hat{G}(\omega'-\overline{\omega})\hat{V}_{s,-}\hat{\rho}_{\mathrm{mat},a}\right\}\mathrm{tr}\left\{\hat{V}_{s,+}\hat{G}(\overline{\omega}-\omega')\hat{V}_{z,-}\hat{G}(-\omega')\hat{V}_{s,-}\hat{\rho}_{\mathrm{mat},b}\right\}.
\end{align*}
In order to compare $S^\mathrm{(\mathrm{casc})}_4$ to the classical contribution, we can employ the substitutions $\overline{\omega}=\omega'-\Omega $ if $\omega'-\overline{\omega}\geq 0$ and $\overline{\omega}=\omega'+\Omega $ if $\omega'-\overline{\omega}\leq 0$ to obtain
\begin{align*}
S^\mathrm{(\mathrm{casc})}_4&=\left(\frac{N\omega_\mathrm{p}L}{c_0}\right)^2\int_0^\infty \mathrm{d}\omega\mathrm{d}\omega'\mathrm{d}\Omega\mathbb{E}(\omega,\omega',\omega+\Omega,\omega'+\Omega)\left(\frac{\hbar\Omega}{2C}+i\frac{\hbar c_0}{LC}\right)\\
&\qquad\times\mathrm{tr}\left\{\hat{V}_{s,+}\hat{G}(\omega)\hat{V}_{z,-}\hat{G}(-\Omega)\hat{V}_{s,-}\hat{\rho}_{\mathrm{mat},a}\right\}\mathrm{tr}\left\{\hat{V}_{s,+}\hat{G}(\Omega)\hat{V}_{z,-}\hat{G}(-\omega')\hat{V}_{s,-}\hat{\rho}_{\mathrm{mat},b}\right\}\\
&\quad-\left(\frac{N\omega_\mathrm{p}L}{c_0}\right)^2\int_0^\infty \mathrm{d}\omega\mathrm{d}\omega'\mathrm{d}\Omega\mathbb{E}(\omega,\omega',\omega-\Omega,\omega'-\Omega)\left(\frac{\hbar\Omega}{2C}-i\frac{\hbar c_0}{LC}\right)\\
&\qquad\qquad\times\mathrm{tr}\left\{\hat{V}_{s,+}\hat{G}(\omega)\hat{V}_{z,-}\hat{G}(\Omega)\hat{V}_{s,-}\hat{\rho}_{\mathrm{mat},a}\right\}\mathrm{tr}\left\{\hat{V}_{s,+}\hat{G}(-\Omega)\hat{V}_{z,-}\hat{G}(-\omega')\hat{V}_{s,-}\hat{\rho}_{\mathrm{mat},b}\right\}.
\end{align*}

In the end, the total normally-ordered second moment $\Gamma$ is given by
\begin{align}
\begin{split}
\Gamma&=\mathrm{tr}\left\{\left[\frac{N\omega_\mathsf{p}L}{c_0 }\int_0^\infty\!\!\mathrm{d}\Omega~(\hat{E}_{\mathrm{THz},s}(\Omega)D(\Omega,\theta)+H.c.)\right]^2\hat{\rho}_\mathrm{THz}\right\}\\
&\qquad +\left(\frac{N\omega_\mathsf{p}L}{c_0 }\right)^2\int_0^\infty\!\!\mathrm{d}\Omega~\Big(\Re\left\{D(\Omega,\theta)\left[D_\mathrm{q}(\Omega,\theta)+D_\mathrm{casc}(\Omega,\theta)\right]\right\}\\
&\qquad\qquad -2\frac{\hbar c_0}{CL}\Im\left\{D(\Omega,\theta)\left[D_\mathrm{q}\Omega,\theta)+D_\mathrm{casc}(\Omega,\theta)\right]\right\}\Big).\label{Gammaarbf}
\end{split}
\end{align}
The susceptibilities contained in the detection windows are given by
\begin{align}
\begin{split}
\chi^{(2)}_{+rs}(-(\omega_2+\omega_1);\omega_2,\omega_1)&=\frac{2^{-1-[\mathrm{sgn}(r)+\mathrm{sgn}(s)]/2}}{\varepsilon_0\hbar^2}\sum_{a,b=g',f} \mu_{gb}\mu_{ba}\mu_{ag} \\
&\qquad\times\left[I_{bg}(\omega_2+\omega_1)I_{ag}(\omega_1)+\mathrm{sgn}(s)I_{ab}(\omega_2+\omega_1)I_{gb}(\omega_1)\right.\\
&\qquad\qquad+\mathrm{sgn}(r)\mathrm{sgn}(s)I_{ga}(\omega_2+\omega_1)I_{gb}(\omega_1)\\
&\qquad\qquad+\left.\mathrm{sgn}(r)I_{ab}(\omega_2+\omega_1)I_{ag}(\omega_1)\right],
\end{split}
\end{align}
with $I_{ab}(\omega)=\frac{1}{\omega-\omega_{ab}+i\gamma_{ab}}$.
Note that only the classical contribution given in the first line of Eq.~\eqref{Gammaarbf} actually depends on the state of the THz field. Furthermore, the expression inside of the square brackets of the classical contribution represents the first correction according to the THz field to the mean value of the electro-optic signal $\hat{\mathcal{S}}(\theta)$ itself.

Finally, Eq.~\eqref{gammaI} can be obtained by evaluating Eq.~\eqref{Gammaarbf} for $\hat{\rho}_\mathrm{THz}=\Ket{0}\!\Bra{0}$.

%

\end{document}